\documentclass[aps,prl,fleqn,twocolumn,superscriptaddress,floatfix]{revtex4}

\usepackage{graphicx}
\usepackage{amsmath,amssymb,amsfonts}

\begin{document}


\title{Azimuthal Asymmetry of Direct Photons in High Energy Nuclear Collisions}

\author{S.~Turbide}
\author{C.~Gale}
\affiliation{Department of Physics, McGill University, Montr\'eal, 
  Canada H3A 2T8}

\author{R.~J.~Fries}
\affiliation{School of Physics and Astronomy, University of Minnesota,
             Minneapolis, MN 55455}

\date{\today}

\begin{abstract}
We show that a sizeable azimuthal asymmetry, characterized by a 
coefficient $v_2$, is to be expected
for large $p_T$ direct photons produced in non-central high energy nuclear collisions. 
This signal is generated by photons radiated by jets interacting with
the surrounding hot plasma. The anisotropy is out of phase by an angle
$\pi/2$ with respect to that associated with the elliptic anisotropy of 
hadrons, leading to negative values
of $v_2$. Such an asymmetry, if observed,
could be a signature for the presence of a quark gluon plasma and
would establish the importance of jet-plasma interactions as a source of
electromagnetic radiation. 
\end{abstract}

\maketitle

At the Relativistic
Heavy Ion Collider (RHIC) and soon at the Large Hadron Collider (LHC), nuclei
are collided at ultrarelativistic energies in order to create a new state of
matter: the  
quark gluon plasma (QGP) \cite{Harris:1996zx}. Emitted particles that only 
interact through the electroweak 
interactions are very unlikely to interact again 
despite the dense medium which is created. Thus they are able to carry 
to the detectors 
information about the state of the system at the time they were created 
\cite{Feinb:76}.
Photons and leptons thus constitute a unique class of penetrating probes. 

The two most interesting sources of photons are those where the 
plasma is directly involved in the emission. These are the thermal 
radiation from the hot QGP \cite{KaLiSei:91} and the radiation
induced by the passage of high energy jets through the plasma 
\cite{FMS:02,Zakharov:04,TGJM:05}. The thermal radiation is emitted
predominantly with low transverse momentum $p_T$ and has to compete
with photon emission from the hot hadronic gas at later times 
\cite{XSB:92,Turb:04}. Photons from jets are an important source at 
intermediate $p_T$, where they compete with photons from 
primary hard scatterings between partons of the nuclei \cite{Owens:86}. 
They probe the thickness of the medium: the longer the path of the jet, 
the more photons are emitted.

Obviously, measuring photons from either of the sources involving the
quark gluon plasma would be an
important step toward establishing its existence and would provide a stringent
test of the reaction dynamics. 
We refer the reader to
\cite{TGJM:05} for a discussion of the different photon sources.
Recently, the PHENIX collaboration at RHIC published their first results
on direct photons measured in Au+Au collisions at $\sqrt{s_{\rm{NN}}}=200$ GeV 
\cite{phenix:05gammaauau}. Note that direct photons here are defined as the 
total inclusive photon yield minus the photons originating from the decay of 
hadrons like $\pi^0$ and 
$\eta$. The sources discussed above will contribute to the direct photon 
signal. The data covers
the intermediate and high-$p_T$ range where thermal photons
are not a leading source. This makes it particularly attractive
to look for photons from jet-plasma interactions.

Nuclear collisions at finite impact parameter $b>0$ start out in an initial
state which is not azimuthally symmetric around the beam axis. Instead,
the initial overlap zone of the two nuclei has an ``almond'' shape. 
Therefore particle spectra measured in the final state are not necessarily
isotropic around the beam axis. It has been
argued that the translation of the original space-time asymmetry into
a momentum space anisotropy can reveal important information about the
system \cite{Olli:92}. Two different mechanisms are important here:
hydrodynamic pressure for the bulk of the matter, at low- to 
intermediate-$p_T$, and a simple optical-depth argument for intermediate- 
to high-$p_T$ particles. We introduce a novel third 
mechanism in this Letter.

Let us define the reaction plane as the plane spanned by the beam axis and the
impact parameter of the colliding nuclei. 
For the bulk of the matter, the initial space-time asymmetry leads to an anisotropic 
pressure gradient which is larger where the material is thinner, 
i.e. in the event plane.
This translates into a larger flow of matter in this direction.
The anisotropy is analyzed in terms of Fourier coefficients 
$v_k$ defined from the particle yield $dN/p_T dp_Td\phi$ as~\cite{VoloZhang:94}
\begin{equation}
  \frac{dN}{p_Tdp_Td\phi} = \frac{dN}{2\pi p_Tdp_T} \left[ 1+
  \sum_k 2 v_k(p_T) \cos( k\phi) \right]
\end{equation}
where the angle $\phi$ is defined with respect to the event plane.
At midrapidity all odd coefficients vanish for symmetry reasons, leaving
the coefficient $v_2$ to be the most important one. Its size determines the
ellipsoidal shape of the anisotropy. It is clear that the elliptic asymmetry 
coefficient $v_2$ is always positive for hadrons at low and intermediate
$p_T$ due to the hydrodynamic flow.
On the other hand, jets lose more energy when they are born into a
direction where the medium is thicker, i.e.\ out of the reaction plane. 
The stronger jet quenching leads to fewer hadrons at intermediate and 
high $p_T$ emitted into this direction. This ``optical $v_2$'' 
is not associated with flow but with absorption and implies positive $v_2$
for hadrons from jets. 
Measurements at RHIC for several hadron species confirm this 
behavior~\cite{star:02v2}.

In this Letter, we discuss $v_2$ of direct photons. We 
concentrate on intermediate and high $p_T$ and the $v_2$ from all the relevant
processes.
We define a mechanism that works by absorption of particles or jets going 
through the medium as optical. It turns out that in some cases
a new inverse-optical mechanism is in place for photons: there are more 
of them emitted into the direction where the nuclear overlap zone is 
thicker, thus leading to a situation where the anisotropy is shifted by 
a phase $\pi/2$. Correspondingly $v_2$ is negative in this case.

Let us now discuss the different contributions to the direct photon spectrum.
Direct photons from primary hard Compton and annihilation processes 
$a+b\to \gamma+c$ are produced symmetrically with
\begin{equation}
\label{N-N}
  \frac{dN^{\rm N-N}}{p_T dp_Td\phi} = T_{AB} f_{a/A} \otimes
  \sigma_{a+b\to \gamma+c} \otimes f_{b/B}.
\end{equation}
Here $\sigma_{a+b\to \gamma+c}$ is the cross section between partons,
$f_{a/A}$, $f_{b/B}$ are parton distribution functions in the nuclei 
$A$ and $B$ and $T_{AB}$ is the overlap factor of the nuclei.
The primary hard direct photons do not suffer any final state effect and
do not exhibit any elliptic asymmetry.

Jets from processes ($a+b \to c+d$) are also produced symmetrically, however 
they are quenched once they start to propagate through the plasma. 
This is the optical mechanism that leads to positive $v_2$ for hadrons 
fragmenting from jets. We expect photons fragmenting from such jets 
in the vacuum ($c\to c+\gamma$, after $c$ propagated through the medium) to
exhibit the same anisotropy. Their yield at midrapidity is given by
\begin{equation}
  \frac{dN^{\rm jet-frag}}{p_T dp_Td\phi} = \sum_f \left. 
  \frac{dN^f(\phi)}{dq} \right|_{q=p_T/z}  \otimes D_{f/\gamma}(z,p_T)
\end{equation}
where $dN^f(\phi)/dq$ is the distribution of jet partons $f$ with
momentum $q$ traveling into the direction given by the angle $\phi$,
and $D_{f/\gamma}$ is the photon fragmentation function.

The interaction of jets with the medium can also produce photons in 
different ways: (i) scattering off plasma components can induce photon
bremsstrahlung, (ii) hard leading partons may annihilate with thermal ones
($q+\bar q \to \gamma+g)$, 
or they can participate in Compton scattering ($q (\bar{q}) +g \to q (\bar{q})
+\gamma$). 
The latter case is also called jet-photon conversion, because
the cross section is dominated by transfer of the entire jet momentum to 
the photon, $\mathbf{p}_\gamma\approx \mathbf{p}_{\rm jet}$.  
The jet-photon conversion yield at midrapidity is given by~\cite{wong}
\begin{multline}
  \frac{dN^{\rm jet-th}}{p_T dp_T d\phi dy} = \int d^4 x
  \frac{\alpha\alpha_s T^2 }{8\pi^2} \sum_q \left( \frac{e_q}{e} \right)^2
  \\ \times f_q(x;p_T,\phi,y) \left[ 2 \ln\frac{4E_\gamma T}{m^2} -C \right]
\end{multline}
with $C=2.332$ and $m^2=4\pi\alpha_s T^2/3$. The distribution of jet partons
$f_q(x;p_T,\phi,y)$ at a space-time point $x$ is determined from the
time dependent spectrum of jet partons propagating in the plasma, 
$dN^q/dq(\tau)$, as discussed in \cite{FMS:02,TGJM:05}.
The time dependence is governed by the energy loss through induced gluon
radiation, obtained with the complete leading order
description by Arnold, Moore and Yaffe (AMY) \cite{AMY}.
It is clear that an anisotropy in $\phi$ is introduced by the different
path lengths for jets traveling in and out of the event 
plane, leading to an increased probability for a jet-photon conversion
in the direction where the medium is thicker. Such an inverse optical
effect has not been observed before.

Medium induced bremsstrahlung ($q\to\gamma+q$) is implemented directly in the AMY formalism
through splitting functions $d\Gamma^{q\to q\gamma}/dkdt$. The photon yield from this process is obtained by
\begin{equation}
\label{jet-brem}
  \frac{dN^{\rm jet-br}}{dp_Td\phi} = \int d^2 r_\perp
  \mathcal{P}(\mathbf{r}_\perp)\int^{d}_0 dt dk\frac{dN^q}{dq}\frac{d\Gamma^{q\to q\gamma}(q,k)}{dkdt} 
 \end{equation}
Here $\mathcal{P}(\mathbf{r}_\perp)$ is the spatial distribution of hard
processes creating jets in the transverse plane and $d=d(\mathbf{r}_\perp,\phi)$
is the distance the jet has to travel from $\mathbf{r}_\perp$ into the 
direction of the angle $\phi$ to leave the fireball.
Again it is obvious that the probability for induced bremsstrahlung
to occur increases with the path length $d$ of the jet. Hence these
photons are preferentially emitted into the direction where 
the medium is thicker, leading to negative $v_2$.

Finally, thermal photon emission constitutes another contribution
to the direct photon spectrum.  Thermal photons emitted by the medium are
not prone to any optical effects, but the emitting matter experiences
the anisotropic hydrodynamic push. 
However, the emission of intermediate and large $p_T$ photons peaks strongly 
at very early times where the temperature is highest \cite{kolb}.
We verified that at the time the system generates significant radial and 
elliptic flow the rate of thermal photons at intermediate and high $p_T$
is negligible. 

Let us summarize what we have so far. We identified two processes,
induced bremsstrahlung from jets and jet-photon conversion, that we expect
to exhibit an inverse optical anisotropy ($v_2 <0$). Photons from 
fragmentation 
show the regular optical anisotropy ($v_2>0$), while primary hard and thermal 
photons do not contribute to $v_2$ at intermediate and high $p_T$.

To quantify our arguments, we carry out a numerical calculation for Au+Au 
collisions at RHIC ($\sqrt{s}=200$ GeV).
Photon spectra at midrapidity with their dependence on the
azimuthal angle $\phi$ are calculated as described above for three
different centrality classes. Our modeling of the nuclear collision is 
introduced in \cite{TGJM:05}.
The initial conditions are constrained by 
$\tau_iT_i^3\sim dN(b)/dy/A_\perp(b)$, where the charged particle 
pseudorapidity densities $dN/dy$ can be found in \cite{phobos} and 
$A_\perp$ is the overlap area of the nuclei in the transverse plane.  
With fixed initial time $\tau_i=0.26$ fm/c, the initial temperatures are 
$T_i$=370, 360 and 310 MeV for
centrality classes $0-20\%, 20-40\%$ and $40-60\%$ respectively. 
Comparing with measured photon spectra, a good agreement is obtained, for all centrality classes.  Details will appear elsewhere~\cite{santa_fe}. The coefficients $v_2$ can then be calculated by using
\begin{equation}
  \label{eq:v2}
  v_2 (p_T) = \frac{\int d\phi \cos(2\phi) dN/dp_Td\phi}{
  dN/dp_T}.
\end{equation}

\begin{figure}
  \begin{center}
  \includegraphics*[width=\columnwidth]{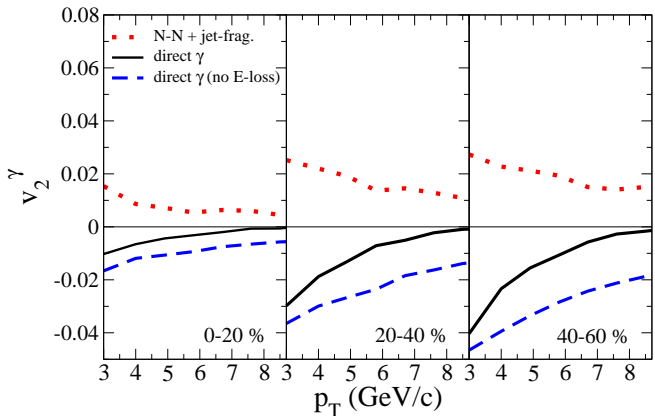}
  \end{center}
  \caption{\label{fig:1} (Color online) Photon $v_2$ as a function of $p_T$ 
for Au+Au collisions at RHIC. Three different centrality bins are given.
  The dotted lines show $v_2$ for primary hard photons and jet fragmentation only,
  and the solid lines show all direct photons. Energy loss is included in both cases.
  The dashed line is the same as the solid line but without energy loss
  of jets taken into account.}
\end{figure}

Fig.\ \ref{fig:1} shows the coefficient $v_2$ as a function of
$p_T$ for Au+Au collisions at RHIC and for the centrality classes 0-20\%,
20-40\% and 40-60\%. The dotted lines give the results for primary hard direct
photons and photon fragmentation. As expected photons from fragmentation lead to 
a positive $v_2$ which is diluted by adding primary hard photons
The solid lines are the results when bremsstrahlung, 
jet-photon conversion and thermal photons are included. They meet our
expectations for $v_2$ of direct photons including all source
discussed above.
The $v_2$ for induced bremsstrahlung and jet-photon conversion is 
indeed negative. Together they are able to overcome the positive $v_2$ from
fragmentation, leading to an overall negative elliptic asymmetry for direct
photons at moderate $p_T$. 
Only above 8 GeV/$c$ the direct photon $v_2$ is again positive, 
because the yield of photons from fragmentation is dominating over medium
induced bremsstrahlung \cite{TGJM:05} in this range.
The dashed lines in Fig.\ \ref{fig:1} show the $v_2$ for direct photons
with no jet energy loss included. In this case, fragmentation photons do not
exhibit an anisotropy and the elliptic asymmetry is only due to jet-photon 
conversion.  Measurements of $v_2$ with sufficient accuracy could therefore
constrain models for jet energy loss.
The absolute size of $v_2$ is not large, about 2-3\%
for the 20-40\% centrality bin around $p_T=4$ GeV/$c$ and up to 5\%
for the more peripheral bin. The reason is that the signal is diluted by 
isotropic photons (primary hard and thermal)
and partially cancelled between the optical and inverse optical mechanisms. 
We also checked the dependence of $v_2$ on the temperature of
the medium, by varying the initial temperature $T_i$ with 
$\tau_i T_i^3$ kept constant. The resulting changes are small: a change in $T_i$ of 40$\%$ generates a shift in $v_2$ of less than 20$\%$.

\begin{figure}
  \begin{center}
  \includegraphics*[width=\columnwidth]{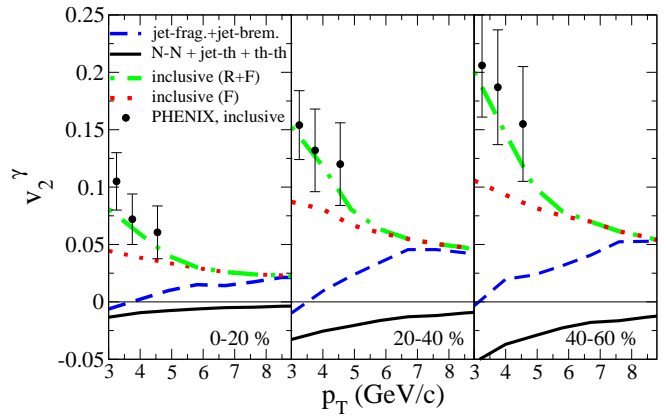}
  \end{center}
  \caption{\label{fig:2} (Color online) $v_2$ as a function of $p_T$ for Au+Au collisions at RHIC. 
 The dashed line shows jet-fragmentation and induced bremsstrahlung only while
 the solid lines give jet-photon conversion, primary hard and thermal photons.
 The dotted line shows direct photons and the background from decay of 
 neutral mesons coming from jets. The dot-dashed line adds photons 
 from decay of recombined pions as well and can be compared to the inclusive
 photon $v_2$ measured by PHENIX \cite{Adler:2005rg}.}
\end{figure}

In Fig.\ \ref{fig:2} we
show some $v_2$ signals that should be detectable at RHIC in the near
future. The dotted lines show $v_2$ for inclusive photons before
background subtraction. In this case the $v_2$ signal 
is dominated by contributions from decaying $\pi^0$ and $\eta$ hadrons. The 
resulting $v_2$ is positive and larger in magnitude. Only hadrons from
fragmentation have been included. However, it has been shown that hadron
production up to a $p_T$ of 4 to
6 GeV/$c$ receives significant 
contributions from recombination of quarks \cite{FMNB:03}.
The dot-dashed lines show the $v_2$ of inclusive 
photons if decays of $\pi^0$ and $\eta$ from recombination are included, 
using \cite{FMNB:03} with parameters consistent with our jet fragmentation
calculation. Data on $v_2$ for inclusive 
photons have been measured by the PHENIX collaboration
\cite{Adler:2005rg}. 
Our calculations for the total inclusive photons including decays from
recombined hadrons agrees well with their results.

A very exciting option for the future is the possibility to 
experimentally distinguish between direct photons associated with jets and 
isolated direct photons. Photons from jet fragmentation and induced 
bremsstrahlung are in the former category, while the latter includes 
primary hard and thermal photons, and photons from jet-photon conversion.
The dashed line in Fig.\ \ref{fig:2} shows the result for the fragmentation
and bremsstrahlung processes only. They contribute with different signs and
one notes a characteristic change of sign from negative values at low $p_T$
to larger positive values, up to 5\%, at large $p_T$, where fragmentation
dominates.
The solid line shows the $v_2$ of all isolated direct photon processes,
including primary hard direct photon, thermal and jet-photon conversion. 
Jet-photon conversion is the only source of an anisotropy, so that 
the resulting $v_2$ is relatively large and negative.

We briefly mention here that the inverse optical mechanism for photon 
$v_2$ could be also at work for jets going through hadronic matter.
Possible examples are hadronic processes after hadronization of the jet, 
e.g. $\rho+\pi \to \gamma+\pi$, as well as Compton, annihilation and
bremsstrahlung processes with partons in surrounding hadrons before 
hadronization. In the first case the $p_T$ of the interacting jet hadron
will be smaller than the original $p_T$ of the jet due to energy loss and
hadronization, shifting the emitted photons to smaller $p_T$. In the second
case the yield will ultimately depend on the unknown parton content of the hot
hadron phase. However the Compton and annihilation yields 
have a $T^2$ dependence at intermediate $p_T$, hence any 
reasonable assumption of the parton content and the temperature of the
hadronic phase will lead to small yields at intermediate $p_T$. The admixture
of photon $v_2$ from jet-hadron interactions is therefore suppressed. 

To summarize, we present the first calculation of the lowest order azimuthal
asymmetry coefficient $v_2$ for direct photons in high energy
 nuclear collisions.
Jets interacting with a deconfined quark gluon plasma provide photons 
exhibiting an inverse optical anisotropy with characteristic negative values
of $v_2$. An experimental confirmation would emphasize the existence
of a quark gluon plasma and confirm jet-medium interactions as important
sources of photons at intermediate $p_T$. 
The $v_2$ signal is generally of order 3-5\% and should be experimentally 
accessible at RHIC. Even more promising would be a 
separation of direct photons emitted in a jet from isolated photons. 
Both sources carry their own characteristic $p_T$ dependence for $v_2$.
The arguments presented here for direct photons immediately apply to 
production of lepton pairs as well. Dileptons from annihilation of
jets in the medium and from medium-induced virtual photon bremsstrahlung
\cite{SGF:02} should also exhibit negative $v_2$.

\begin{acknowledgments} We thank B.\ Cole  and S.\ Jeon for stimulating 
discussions, and G.\ Moore for his help with details of the AMY energy loss. 
R.\ J.\ F.\ thanks the nuclear theory group at McGill University for their 
hospitality while part of this work was carried out.
This work was supported in part by the Natural Sciences and Engineering 
Research Council of Canada, by le Fonds Nature et Technologies du Qu\'ebec,
and by DOE grant DE-FG02-87ER40328.
\end{acknowledgments}

\end{document}